\documentclass{l4dc2027}

\title[Sample-Efficient Model Predictive Control via Proximal Policy Optimization]{Sample-Efficient Model Predictive Control via Proximal Policy Optimization}
\usepackage{times}
\usepackage{booktabs}
\usepackage{caption}

\author{%
 \Name{Jiarui Wang} \Email{jwang486@jhu.edu}\\
 \Name{Sina Sharifi} \Email{sshari12@jhu.edu}\\
 \Name{Mahyar Fazlyab} \Email{mahyarfazlyab@jhu.edu}\\
 \addr Department of Electrical and Computer Engineering,\\
 Johns Hopkins University, Baltimore, MD 21218, USA
}

\newcommand{\E}{\mathbb{E}}

\newcommand{\KL}{D_{\mathrm{KL}}}

\usepackage{tikz}
\usepackage{pgfplots}
\pgfplotsset{compat=1.18}

\newtheorem{assumption}{Assumption}

\usepackage{enumitem}
\usepackage{float}

\begin{document}

\maketitle

\begin{abstract}%
Sampling-based model predictive control (MPC) enables
real-time control of complex nonlinear systems but
often incurs substantial computational costs due to
repeated trajectory simulations.
In this paper, we bridge policy optimization in
reinforcement learning and sampling-based MPC to
improve computational efficiency through sample reuse.
Starting from a KL-regularized distributional formulation
of trajectory optimization, we derive an exact
performance-difference identity based on a signed
exponential advantage function.
This identity motivates two algorithms,
PPO-MPPI and SPO-MPPI, which adapt proximal policy
optimization techniques to sampling-based MPC,
enabling multiple optimization updates using
previously evaluated trajectories without additional
forward simulations.
We establish population-level monotonic improvement
guarantees for both methods and prove convergence
to first-order stationarity for PPO-MPPI under
fixed-covariance Gaussian sampling distributions.
Experiments on quadruped velocity tracking and
quadrotor path tracking demonstrate that the proposed
methods achieve competitive control performance
at substantially lower computational cost than
multi-stage sampling approaches.
Our results establish a principled connection between
policy optimization and sampling-based MPC and
highlight the potential of trajectory reuse for
sample-efficient predictive control.
\end{abstract}


\section{Introduction}
\label{sec:intro}
Sampling-based model predictive control (MPC), such as model predictive path integral (MPPI) control \citep{williams2017information} and the cross-entropy method (CEM) \citep{botev2013cross}, has emerged as a popular framework for real-time control of complex robotic systems \citep{alvarez2025real, xue2025full}. These methods operate by generating candidate control sequences, evaluating their performance through forward simulation of the system dynamics, and using the resulting trajectory costs to update control actions. Their ability to handle nonlinear, nonsmooth, and black-box dynamics without requiring explicit derivatives, together with their natural parallelizability on GPUs, has enabled successful applications across a wide range of robotic control tasks~\citep{honda2026model}. Despite these advantages, sampling-based MPC can be computationally demanding, as evaluating large numbers of candidate trajectories often requires expensive forward simulations, particularly for high-fidelity simulators involving contact dynamics, collision checking, or learned dynamics models. 

A closely related challenge arises in reinforcement learning (RL), where collecting trajectories through interactions with the environment can be prohibitively expensive. To improve sample efficiency, modern policy optimization algorithms seek to extract more learning progress from previously collected trajectories. Notably, proximal policy optimization (PPO) \citep{schulman2017proximal} and simple policy optimization (SPO) \citep{xie2024simple} employ likelihood-ratio-based objectives that enable multiple policy updates using the same batch of experience, while incorporating mechanisms to mitigate excessive distribution shift. These advances suggest a natural opportunity to bridge policy optimization in RL and sampling-based MPC: \textit{Can we leverage sample-reuse techniques from RL to improve the computational efficiency of sampling-based MPC without sacrificing control performance?}

In this paper, we bridge these two paradigms by developing a policy optimization framework for sample-efficient MPC. Our starting point is a distributional formulation of trajectory optimization, in which a general parameterized distribution over open-loop control sequences is optimized by minimizing a KL-regularized free-energy objective \citep{fazlyab2026model}. By expressing this objective under a fixed sampling distribution through a change of measure, we enable multiple optimization steps using \textit{the same} batch of evaluated trajectories, without requiring additional forward simulations. Building on this formulation, we introduce two complementary update mechanisms inspired by modern policy optimization: PPO-MPPI, which employs likelihood-ratio clipping, and SPO-MPPI, which incorporates regularization to mitigate distribution shift. Both methods operate within the same optimization framework and are implemented in a receding-horizon fashion, combining sample reuse with online feedback control. The main contributions of this paper are summarized
as follows:

\begin{itemize}[leftmargin=*]
    \item \textbf{Free-energy performance-difference identity.}
    Inspired by the classical performance-difference
    identity in RL~\citep{kakade2002approximately},
    we derive an exact analogue for sampling-based MPC
    by defining a signed exponential advantage function.
    This identity establishes a direct connection
    between free-energy minimization and
    advantage-based policy optimization, providing
    the theoretical foundation for sample reuse.

    \item \textbf{Proximal policy optimization for MPC.}
    We develop PPO-MPPI and SPO-MPPI, which enable
    multiple optimization updates using previously
    evaluated trajectories through likelihood-ratio
    objectives. We establish population-level
    monotonic improvement guarantees for both methods
    and prove convergence to first-order stationarity
    for PPO-MPPI under fixed-covariance Gaussian
    sampling distributions.

    \item \textbf{Efficient robotic control through sample reuse.}
    We evaluate the proposed methods on quadruped
    velocity tracking and quadrotor path tracking.
    The results demonstrate competitive control
    performance with substantially lower computational
    cost than multi-stage sampling approaches,
    highlighting the benefits of reusing trajectories
    rather than repeatedly simulating new ones.
\end{itemize}

\subsection{Related Work}

Sampling-based MPC has been studied from several
perspectives, including control-as-inference
\citep{levine2018reinforcement,honda2026model, wang2026generalized},
KL-regularized optimization
\citep{williams2015model,williams2017information},
and mirror descent
\citep{miyashita2018mirror,okada2018acceleration}.
Particularly relevant to our work,
\citet{fazlyab2026model} derive a reduced
free-energy objective over general parameterized
sampling distributions and show that classical
MPPI, under a fixed-covariance Gaussian
parameterization, corresponds to a preconditioned
gradient update. We build on this formulation
to develop a sample-reusing optimization framework
for general parameterized control distributions.

A central challenge in sampling-based MPC is the
computational cost of trajectory evaluation, which
has motivated methods for improving sample
efficiency. The improved cross-entropy method
(iCEM) reuses promising trajectories across
optimization iterations \citep{pinneri2021sample},
while importance-sampling-based methods reuse
previously evaluated candidates in black-box
optimization
\citep{pourchot2018importance,shirakawa2018sample}.
However, the effectiveness of sample reuse
depends on the discrepancy between the updated distribution and the distribution
that generated the samples.
Recent work by \citet{walker2026sampling}
develops a trust-region stochastic search
framework for MPC that minimizes expected
trajectory cost subject to KL-divergence
and optional entropy constraints.
A similar concern arises in policy optimization
for reinforcement learning, where trust-region
methods constrain policy updates to limit
distribution shift
\citep{schulman2015trust,achiam2017constrained}.
PPO \citep{schulman2017proximal} and SPO
\citep{xie2024simple} further enable multiple
updates using the same batch of experience
through likelihood-ratio clipping and
regularization, respectively.
Motivated by these developments, we directly
optimize the reduced free-energy objective
using previously evaluated trajectories,
incorporating PPO- and SPO-inspired mechanisms
to mitigate distribution shift during
repeated updates.

\section{Background and Problem Formulation}
\label{sec:problem}

Consider the discrete-time dynamical system
$x_{t+1}=f(x_t,u_t)$ and the finite-horizon trajectory
optimization problem
\begin{equation}
    J^\star(x)
    =
    \inf_{U\in\mathcal U} J(U;x),
    \qquad
    J(U;x)
    =
    \sum_{t=0}^{H-1}\ell(x_t,u_t)
    +\ell_f(x_H),
    \label{eq:trajectory_problem}
\end{equation}
where $x_0=x$, $U=(u_0,\ldots,u_{H-1})$, and
$\mathcal U$ denotes the set of admissible control
sequences. Throughout each planning solve, the initial
state $x$ is fixed and its dependence is suppressed.
Evaluating $J(U)$ requires a forward rollout of the
dynamics (e.g., a simulator), but neither the dynamics nor the trajectory cost need to be differentiable. We begin by lifting \eqref{eq:trajectory_problem}
from individual control sequences to probability
distributions over $\mathcal U$:
\begin{equation}
    J^\star
    =
    \inf_{\rho\in\mathcal P(\mathcal U)}
    \E_\rho[J(U)].
    \label{eq:lifted_problem}
\end{equation}
Although this reformulation preserves the optimal
value, its solutions may be degenerate and unsuitable for sampling. In particular,
if a minimizer $U^\star$ exists, the Dirac measure
$\delta_{U^\star}$ solves \eqref{eq:lifted_problem}. To prevent degenerate solutions, consider
a parameterized family of proposal distributions
$\{\pi_\theta:\theta\in\Theta\}$, from which samples
can be drawn and densities evaluated. For a fixed
proposal $\pi_\theta$ and temperature $\tau>0$,
define the variational objective
\begin{equation}
    V(\theta)
    =
    \inf_{\rho\ll\pi_\theta}
    \left\{
        \E_\rho[J(U)]
        +
        \tau\KL(\rho\|\pi_\theta)
    \right\},
    \label{eq:variational}
\end{equation}
where $\rho$ is absolutely continuous with respect to $\pi_\theta$ ($\rho\ll\pi_\theta$).
Here $\pi_\theta$ serves as a tractable reference (or proposal) 
distribution, and  the KL regularizer penalizes excessive
concentration relative to the proposal and balances
the preference for low-cost trajectories against their likelihood under $\pi_\theta$.  Assuming $J$ is measurable and bounded below, the Gibbs variational principle yields the unique
solution of \eqref{eq:variational} as
\begin{equation}
    \rho_\theta^\star(U)
    =
    \frac{
        \pi_\theta(U)e^{-J(U)/\tau}
    }{
        Z(\theta)
    }, \quad  Z(\theta)
    :=
    \E_{\pi_\theta}
    \left[e^{-J(U)/\tau}\right]
    >0.
    \label{eq:gibbs}
\end{equation}
where $Z(\theta)$ is the partition function (see Appendix \ref{app:subsec:gibbs distribution} for a derivation).  Consequently, eliminating the auxiliary distribution
$\rho$ reduces the variational objective, or free energy, to
\begin{equation}
    V(\theta)=
    -\tau\log Z(\theta)=-\tau\log
    \E_{\pi_\theta}
    \left[e^{-J(U)/\tau}\right]
    \label{eq:free_energy_problem}
\end{equation}
Since $\E_\rho[J]\geq J^\star$ and
$\KL(\rho\|\pi_\theta)\geq 0$ for any $\rho \ll \pi_{\theta}$, the variational
objective satisfies
\begin{equation}
    J^\star
    \leq
    V(\theta)
    \leq
    \E_{\pi_\theta}[J(U)],
    \label{eq:free_energy_bounds}
\end{equation}
Minimizing $V(\theta)$ therefore seeks the tightest
upper bound on the optimal trajectory cost within
the chosen proposal family. Moreover, compared with
minimizing the expected cost $\E_{\pi_\theta}[J(U)]$,
the free-energy objective places greater emphasis on
low-cost trajectories through exponential weighting,
rather than treating all sampled costs linearly.

\subsection{Preconditioned Gradient Descent and MPPI}
\label{sec:mppi_gradient}

Suppose the proposal density $\pi_\theta$ is
differentiable in $\theta$, has parameter-independent
support, and satisfies the regularity conditions
needed to interchange differentiation and integration.
Differentiating \eqref{eq:free_energy_problem} yields
\begin{equation}
    \begin{aligned}
    \nabla V(\theta)
    &=
    -\tau
    \frac{
        \E_{\pi_\theta}
        \left[
            e^{-J(U)/\tau}\nabla_\theta\log\pi_\theta(U)
        \right]
    }{
        \E_{\pi_\theta}
        \left[e^{-J(U)/\tau}\right]
    } =
    -\tau
    \E_{\rho_\theta^\star}
    [\nabla_\theta\log\pi_\theta(U)],
    \end{aligned}
    \label{eq:gradient}
\end{equation}
where the second equality follows from the expression
for $\rho_\theta^\star$.
Since this gradient depends only on the proposal
density and evaluated trajectory costs, it does
not require differentiating through the system
dynamics. A general preconditioned gradient step
for Problem~\eqref{eq:free_energy_problem} takes
the form
\begin{equation}
    \theta^+
    =
    \theta-\alpha P\nabla V(\theta),
    \qquad P\succ0,
    \label{eq:preconditioned_update}
\end{equation}
where $\alpha>0$ is the step size. As established in \citet{fazlyab2026model}, the standard
MPPI update is recovered as a special case of
\eqref{eq:preconditioned_update}.
Specifically, consider a Gaussian proposal
$\pi_\mu=\mathcal N(\mu,\Sigma)$ with fixed
covariance $\Sigma\succ0$, where only the
mean $\mu$ is optimized. Since
\(
    \nabla_\mu\log\pi_\mu(U)
    =
    \Sigma^{-1}(U-\mu)
\),  the free-energy gradient becomes
$
    \nabla_\mu V(\mu)
    =
    -\tau\Sigma^{-1}
    \left(
        \E_{\rho_\mu^\star}[U]-\mu
    \right).
$
Choosing the preconditioner $P=\Sigma/\tau$
in \eqref{eq:preconditioned_update} yields
\begin{equation}
    \mu^+
    =
    \mu
    +
    \alpha
    \left(
        \E_{\rho_\mu^\star}[U]-\mu
    \right).
    \label{eq:mppi}
\end{equation}
For $\alpha=1$, this coincides with the
population-level MPPI mean update. Moreover,
the Fisher information matrix of the Gaussian
proposal family is $F(\mu)=\Sigma^{-1}$.
Thus, MPPI can equivalently be interpreted
as a natural-gradient step with step size
$1/\tau$.

Given independent samples $U_i\sim\pi_\mu$,
the corresponding Monte Carlo approximation is
\begin{equation}
    \mu^+
    =
    \mu
    +
    \sum_{i=1}^{N}
    w_i(U_i-\mu),
    \qquad
    w_i
    =
    \frac{
        e^{-J(U_i)/\tau}
    }{
        \sum_{j=1}^{N}e^{-J(U_j)/\tau}
    }.
    \label{eq:sample_mppi}
\end{equation}
Despite its simplicity, conventional
MPPI, and more generally the gradient update rule \eqref{eq:preconditioned_update},  must generate and evaluate a \textit{new}
batch of trajectories after each proposal
update. For high fidelity simulators, trajectory evaluations can
dominate the computational cost of MPC. Therefore,  repeated sampling limits the number of
optimization iterations that can be performed
within a fixed computational budget. This motivates \textit{reusing } previously evaluated
trajectories across multiple proposal updates. The next section develops this formulation
and its resulting optimization algorithm.

\section{Proximal Policy Optimization for Sampling-Based MPC}
\label{sec:method}

In this section, we establish a connection between policy
optimization in reinforcement learning and sampling-based
model predictive control. In reinforcement learning,
the classical performance-difference identity
of \citep{kakade2002approximately} characterizes
the difference in expected returns between two policies
in terms of an advantage function.
Building on this perspective, we derive an analogous
exact performance-difference identity for the
free-energy objective developed in Section~\ref{sec:problem},
using a change of measure and a signed exponential
advantage function. This identity motivates a PPO-style
clipped surrogate that admits a population-level
improvement guarantee while enabling repeated updates
using previously evaluated trajectories.

\subsection{Free-Energy Performance Difference}
\label{subsec:performance_difference}

Recall the free-energy objective \eqref{eq:free_energy_problem}. At iteration $k$, let $\theta_k$ denote the current
parameter and define $\pi_k=\pi_{\theta_k}$.
For any candidate distribution $\pi_\theta$ satisfying
$\pi_\theta\ll\pi_k$, by change of measure, the free-energy objective \eqref{eq:free_energy_problem} can
be expressed as
\begin{equation}
    V(\theta)
    =
    -\tau\log
    \mathbb{E}_{u\sim\pi_k}
    \left[
        r_\theta(u)
        \exp\left(-\frac{J(u)}{\tau}\right)
    \right], \quad \text{with} \quad r_\theta(u)
    :=
    \frac{\pi_\theta(u)}{\pi_k(u)},
    \label{eq:fe_change_measure}
\end{equation}
where $r_\theta(u)$ is the likelihood ratio. This representation is exact at the population level
and allows candidate sampling distributions to be
evaluated using trajectories generated by $\pi_k$. Define the \textit{signed exponential advantage}
\begin{equation}
    A_k(u)
    =
    \frac{\exp(-J(u)/\tau)}{Z(\theta_k)}-1
    =
    \exp\left(
        \frac{V(\theta_k)-J(u)}{\tau}
    \right)-1.
    \label{eq:exp_advantage}
\end{equation}
where the second equality follows from \eqref{eq:free_energy_problem}. By construction,
$
    \mathbb{E}_{\pi_k}[A_k(u)]=0,
    \ 
    A_k(u)>-1.
$
The advantage is positive for trajectories satisfying
$J(u)<V(\theta_k)$ and negative for trajectories
satisfying $J(u)>V(\theta_k)$. Thus, it distinguishes
trajectories that are favorable or unfavorable
relative to the current free energy. The following result characterizes the exact
difference in free energy between two sampling
distributions.

\begin{proposition}[Free-energy performance difference]
\label{prop:performance_difference}
Suppose $\pi_\theta\ll\pi_k$ and the partition
functions are finite and strictly positive.
Then
\begin{equation}
    V(\theta)-V(\theta_k)
    =
    -\tau\log
    \left(
        1+
        \mathbb{E}_{\pi_k}
        \left[
            r_\theta(u)A_k(u)
        \right]
    \right).
    \label{eq:performance_difference}
\end{equation}
Consequently,
\begin{equation}
    V(\theta)\leq V(\theta_k)
    \quad\Longleftrightarrow\quad
    \mathbb{E}_{\pi_k}
    \left[
        r_\theta(u)A_k(u)
    \right]\geq 0.
    \label{eq:improvement_equivalence}
\end{equation}
\end{proposition}
Proposition~\ref{prop:performance_difference}
provides an exact counterpart to the policy
performance-difference results used in reinforcement
learning~\citep{kakade2002approximately}. Unlike the standard discounted-return
setting, where the state-visitation distributions
of two policies generally differ, the free-energy
difference admits an exact representation under
the current sampling distribution. In
particular, minimizing free energy is equivalent
to maximizing the advantage-based objective
\begin{equation}
    L_k(\theta)
    =
    \mathbb{E}_{\pi_k}
    \left[
        r_\theta(u)A_k(u)
    \right].
    \label{eq:unclipped_surrogate}
\end{equation}
Since $L_k(\theta_k)=0$, any candidate satisfying
$L_k(\theta)\geq 0$ guarantees nonincreasing
free energy.

\subsection{A Proximal Surrogate for Free-Energy
Minimization}
\label{subsec:proximal_surrogate}

Although  \eqref{eq:unclipped_surrogate} is exact
at the population level, directly maximizing its
empirical approximation can lead to large changes
in the sampling distribution. In particular,
optimization may increase the likelihood of a few
favorable sampled trajectories while assigning
little probability to other regions of the control
space. Motivated by proximal policy optimization (PPO) \citep{schulman2017proximal},
we introduce a clipped advantage surrogate
\begin{equation}
\begin{aligned}
    L_k^{\mathrm{clip}}(\theta)
    =
    \mathbb{E}_{u\sim\pi_k}
    \Big[
        \min\big\{
            r_\theta(u)A_k(u),
            \operatorname{clip}
            (r_\theta(u),1-\epsilon,1+\epsilon)
            A_k(u)
        \big\}
    \Big],
\end{aligned}
\label{eq:clipped_surrogate}
\end{equation}
where $\epsilon\in(0,1)$ is the clipping
parameter. The signed advantage plays an essential role
in this construction. For trajectories with
$A_k(u)>0$, the surrogate limits the benefit
of increasing their likelihood ratios beyond
$1+\epsilon$. Conversely, for trajectories
with $A_k(u)<0$, it limits the benefit of
decreasing their likelihood ratios below
$1-\epsilon$. Equivalently, the clipped advantage surrogate can be
written as
\begin{equation}
    L_k^{\mathrm{clip}}(\theta)
    =
    \mathbb{E}_{\pi_k}
    \left[
        \ell_\epsilon
        (r_\theta(u),A_k(u))
    \right], \quad \ell_\epsilon(r,A)
    =
    \begin{cases}
        A\min\{r,1+\epsilon\},
        & A\geq 0,\\[2mm]
        A\max\{r,1-\epsilon\},
        & A<0.
    \end{cases}
    \label{eq:clipped_integrand}
\end{equation}
Crucially, clipping is applied to the
advantage-weighted likelihood ratio $r_\theta(u)A_k(u)$ rather than
directly to the free-energy objective. This
construction retains the interpretation of PPO
while preserving a useful relationship to the
original free-energy minimization problem.

\begin{theorem}[Monotonic improvement]
\label{thm:clipped_improvement}
Under the assumptions of
Proposition~\ref{prop:performance_difference},
the clipped surrogate satisfies
\begin{equation}
    L_k^{\mathrm{clip}}(\theta)
    \leq L_k(\theta)
    \label{eq:surrogate_lower_bound}
\end{equation}
for every candidate $\theta$, with equality
at $\theta=\theta_k$. Consequently, any candidate $\theta_{k+1}$
satisfying
\begin{equation}
    L_k^{\mathrm{clip}}(\theta_{k+1})
    \geq
    L_k^{\mathrm{clip}}(\theta_k)=0
    \label{eq:surrogate_improvement}
\end{equation}
guarantees $V(\theta_{k+1})\leq V(\theta_k)$. 
Moreover, whenever
$1+L_k^{\mathrm{clip}}(\theta)>0$,
\begin{equation}
    V(\theta)-V(\theta_k)
    \leq
    -\tau\log
    \left(
        1+L_k^{\mathrm{clip}}(\theta)
    \right).
    \label{eq:clipped_fe_upper_bound}
\end{equation}
\end{theorem}
Theorem~\ref{thm:clipped_improvement}
establishes that the clipped surrogate $L_k^{\mathrm{clip}}(\theta)$ is a
population-level lower bound on the exact
advantage objective. Maximizing this surrogate
therefore provides a sufficient condition for
free-energy improvement. This result highlights a fundamental connection
between policy optimization and sampling-based MPC:
a PPO-style surrogate, constructed using the
exponential-cost advantage, induces a valid
descent mechanism for the free-energy objective.

\subsection{Convergence Analysis}
\label{subsec:convergence}
We next establish convergence guarantees for the proposed
PPO-MPPI method in the population setting. While
Theorem~\ref{thm:clipped_improvement} guarantees
monotonic improvement whenever the clipped surrogate
does not decrease, it does not establish convergence
to a stationary point. We show that, for Gaussian
sampling distributions with fixed covariance, the
clipped surrogate preserves the local first-order
behavior of the free-energy objective. This property
allows us to establish sufficient descent and
convergence to first-order stationarity.
\begin{assumption}
\label{ass:convergence}
The trajectory cost $J(u)$ is measurable,
independent of $\theta$, and bounded:
$J_{\min}\leq J(u)\leq J_{\max}$.
The  clipping parameter $\epsilon\in(0,1)$
remains fixed across iterations.
\end{assumption}
Consider the Gaussian sampling family
$\pi_\theta=\mathcal{N}(\theta,\Sigma)$,
where $\theta\in\mathbb{R}^d$ denotes the mean
control sequence and $\Sigma\succ 0$ is fixed. To analyze the clipped surrogate, introduce the
change of variable $h=\Sigma^{-1/2}(\theta-\theta_k)$ and define
\begin{equation}
    \label{eq:cal_L_def}
    \mathcal{L}_k(h)
    =
    L_k(\theta_k+\Sigma^{1/2}h),
\end{equation}
and similarly define
$\mathcal{L}_k^{\mathrm{clip}}(h)$. We next 
quantify the approximation introduced by clipping.

\begin{lemma}[Clipping error]
\label{lem:clipping_error}
Suppose Assumption~\ref{ass:convergence} holds,
and define $m=e^{-J_{\max}/\tau}, G=e^{-J_{\min}/\tau}$. 
Then, for every $h\in\mathbb{R}^d$,
\begin{equation}
    0\leq
    \mathcal{L}_k(h)
    -
    \mathcal{L}_k^{\mathrm{clip}}(h)
    \leq
    \frac{G}{m\epsilon}
    \left(e^{\|h\|^2}-1\right).
    \label{eq:clipping_error}
\end{equation}
In particular, for any $a>0$ and
$\|h\|\leq a$,
\begin{equation}
    \mathcal{L}_k^{\mathrm{clip}}(h)
    \geq
    \mathcal{L}_k(h)
    -
    \frac{Ge^{a^2}}{m\epsilon}
    \|h\|^2.
    \label{eq:local_clipping_bound}
\end{equation}
\end{lemma}
Lemma~\ref{lem:clipping_error} shows that
clipping introduces at most a \textit{second-order}
error in a neighborhood of the current iterate.
In particular, the clipped and unclipped
surrogates have the same first-order behavior
at $\theta_k$. 
The following theorem uses this local
approximation to establish sufficient descent and convergence to first-order stationarity. Let
$\|v\|_{\Sigma^{-1}} \!:=\! \sqrt{v^\top\Sigma^{-1}v}$
denote the norm induced by $\Sigma^{-1}$.

\begin{theorem}[Convergence of population PPO-MPPI]
\label{thm:ppo_mppi_convergence}
Suppose Assumption~\ref{ass:convergence}
holds.
At each iteration, let
$\theta_{k+1}$ maximize
$L_k^{\mathrm{clip}}(\theta)$ over a set
$\Theta_k$ containing $\left\{
        \theta:
        \|\theta-\theta_k\|_{\Sigma^{-1}}
        \leq a
    \right\}$
for some fixed $a>0$.
Assume that a global maximizer exists. Then there is a constant $c>0$,
independent of $k$, such that
\begin{equation}
    V(\theta_{k+1})
    \leq
    V(\theta_k)
    -
    c\|\nabla V(\theta_k)\|_\Sigma^2.
    \label{eq:convergence_descent}
\end{equation}
Consequently, the free-energy values
$V(\theta_k)$ converge, $\lim_{k\to\infty}
    \|\nabla V(\theta_k)\|=0$, and, for every $K\geq 1$,
\begin{equation}
    \min_{0\leq k<K}
    \|\nabla V(\theta_k)\|_\Sigma^2
    \leq
    \frac{
        V(\theta_0)-J_{\min}
    }{cK}.
    \label{eq:convergence_rate}
\end{equation}
Thus, population PPO-MPPI achieves an
$O(1/K)$ convergence rate in the
minimum squared gradient norm.
\end{theorem}
Theorem~\ref{thm:ppo_mppi_convergence}
establishes convergence to first-order
stationarity for population PPO-MPPI,
despite the nonsmooth clipping operation.
The key observation is that clipping
preserves the first-order behavior of the
exact advantage surrogate, while its
approximation error is locally quadratic.
Consequently, sufficiently improving
updates of the clipped surrogate induce
sufficient descent in the original
free-energy objective.

\begin{remark}[Approximate surrogate maximization]
\label{rem:approximate_convergence}
The global-maximization requirement in
Theorem~\ref{thm:ppo_mppi_convergence}
can be relaxed. The proof only requires
the update to satisfy
\begin{equation}
    L_k^{\mathrm{clip}}(\theta_{k+1})
    \geq
    \frac{\gamma t}{2\tau^2}
    \|\nabla V(\theta_k)\|_\Sigma^2
\end{equation}
for some fixed $\gamma\in(0,1]$.
Under this sufficient-improvement
condition, the same conclusions hold,
with the descent constant replaced by
$\gamma c$. In particular, the analysis
does not require imposing a hard
trust-region constraint on the final
update; the neighborhood of radius $a$
is used only to construct a guaranteed
improving candidate.
\end{remark}

\subsection{Simple Policy Optimization for Sampling-Based MPC}
\label{subsec:spo_mppi}

Although clipping promotes conservative updates,
it does not \textit{explicitly} constrain the likelihood ratio
$r_\theta(u)$. In particular, once the ratio crosses the
clipping threshold in the favorable direction, the
corresponding gradient vanishes, potentially allowing
large distribution shifts. To address this limitation,
we adapt Simple Policy Optimization (SPO)
~\citep{xie2024simple}, which replaces clipping with
a quadratic penalty on likelihood-ratio deviations. Specifically, we define the SPO-MPPI surrogate
\begin{equation}
    L_k^{\mathrm{SPO}}(\theta)
    =
    \mathbb{E}_{\pi_k}
    \left[
        r_\theta(u)A_k(u)
        -
        \frac{|A_k(u)|}{2\epsilon}
        \big(r_\theta(u)-1\big)^2
    \right],
    \label{eq:spo_surrogate}
\end{equation}
where $\epsilon>0$ controls the strength of
regularization. Unlike PPO clipping, the quadratic
penalty provides corrective gradients when likelihood
ratios deviate from unity. For any fixed
$A_k(u)\neq 0$, the integrand is maximized at
$r_\theta(u)=1+\epsilon\operatorname{sign}(A_k(u))$,
favoring conservative probability updates. Moreover, since the penalty is nonnegative,
$L_k^{\mathrm{SPO}}(\theta)\leq L_k(\theta)$,
with equality at $\theta=\theta_k$.
Consequently, by
Proposition~\ref{prop:performance_difference},
any update satisfying
$L_k^{\mathrm{SPO}}(\theta_{k+1})\geq 0$
guarantees $V(\theta_{k+1})\leq V(\theta_k)$. Thus, SPO-MPPI retains
the free-energy improvement guarantee while
more directly discouraging distribution shifts. The complete algorithm is provided in Appendix \ref{app:subsec:complete algo}.

\section{Experiments}
We compare PPO-MPPI and its SPO-MPPI variant against Vanilla-MPPI and DIAL-MPC~\citep{xue2025full} on quadruped velocity tracking and quadrotor path
tracking tasks. All methods optimize spline knots that are interpolated
to generate continuous action trajectories.
Following~\citep{xue2025full}, we apply action-level annealing by increasing the sampling variance for knots
farther along the planning horizon.
DIAL-MPC additionally performs five stages of
trajectory-level annealing, whereas PPO-MPPI and
SPO-MPPI reuse sampled trajectories for multiple
optimization steps.
Additional experimental details are provided in
the appendix.

\subsection{Quadruped Velocity Tracking}
We evaluate all four methods on a velocity-tracking
task with a Unitree Go2 quadruped.
Starting from its nominal standing configuration,
the robot tracks a constant body-frame forward
velocity of 1.5~m/s.
The action $\mathbf{u}\in\mathbb{R}^{12}$
specifies additive corrections to the nominal
joint positions:
$
\mathbf{q}_j^{\mathrm{cmd}}
=
\operatorname{clip}(
\mathbf{q}_j^0+0.25\mathbf{u},
\mathbf{q}_j^{\min},
\mathbf{q}_j^{\max}
),
$
where $\mathbf{q}_j^0$ denotes the nominal
configuration and the bounds retain the central
$95\%$ of each joint's position range.
Joint position commands are converted into torques
using a PD controller.
The reference gait is a trot, with diagonally
opposite feet synchronized and the two diagonal
pairs offset by half a cycle.
The stage cost penalizes gait and velocity tracking
errors, base orientation, angular velocity,
height deviation, joint displacement, and
joint velocity.
\begin{figure}[t]
    \centering

    \subfigure[\small Vanilla-MPPI gait.]{
        \includegraphics[width=0.485\linewidth]{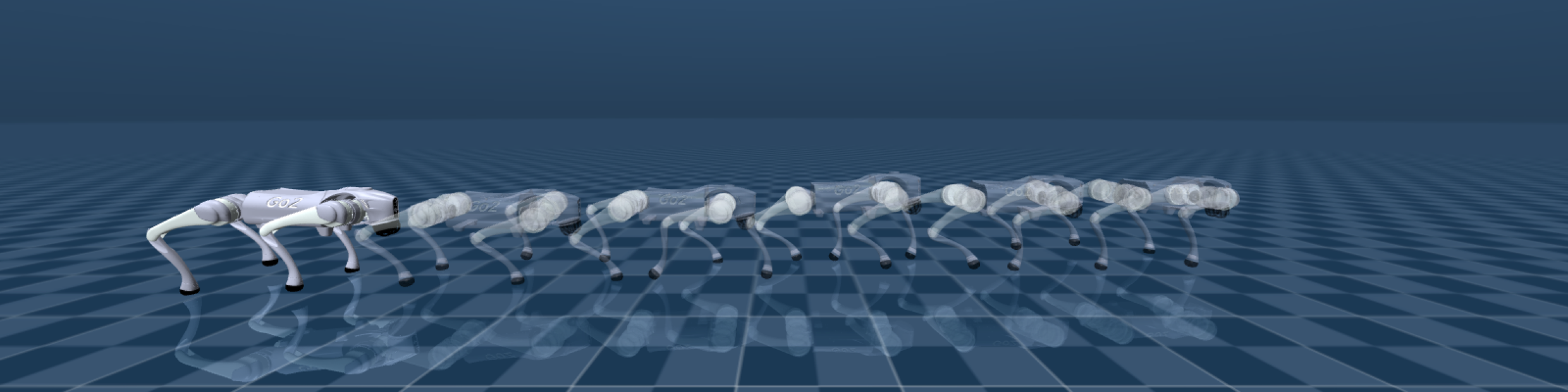}
        \label{fig:go2-walk-gait-naive}
    }%
    \hfill
    \subfigure[\small DIAL-MPC gait.]{
        \includegraphics[width=0.485\linewidth]{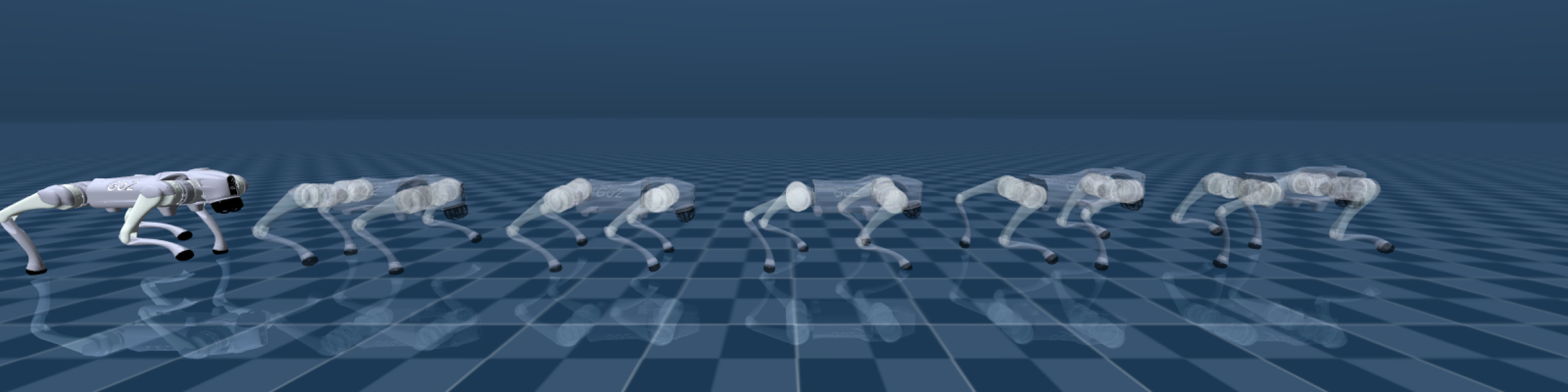}
        \label{fig:go2-walk-gait-dial}
    }

    \par\medskip

    \subfigure[\small PPO-MPPI gait.]{
        \includegraphics[width=0.485\linewidth]{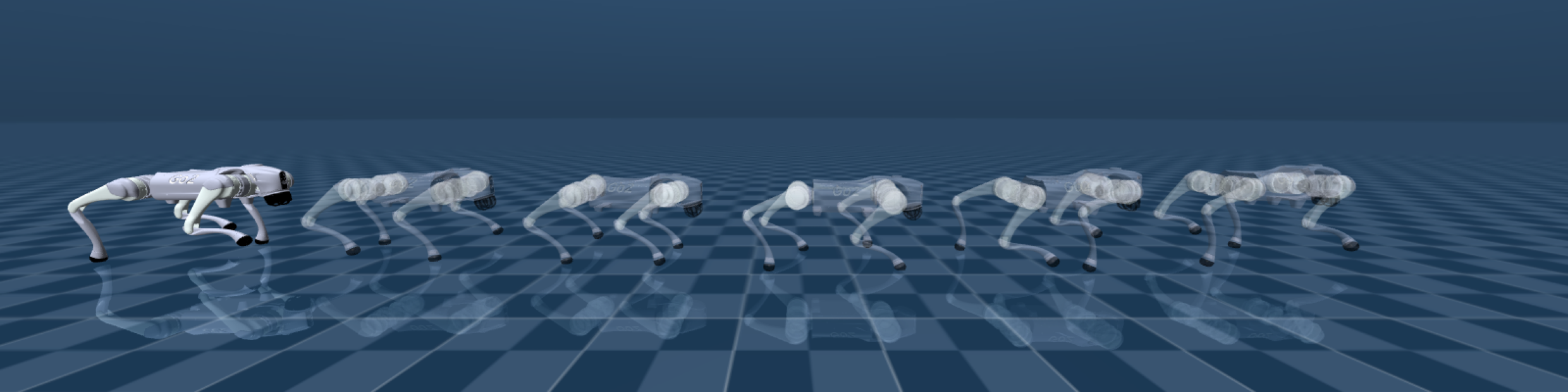}
        \label{fig:go2-walk-gait-ppo}
    }%
    \hfill
    \subfigure[\small SPO-MPPI gait.]{
        \includegraphics[width=0.485\linewidth]{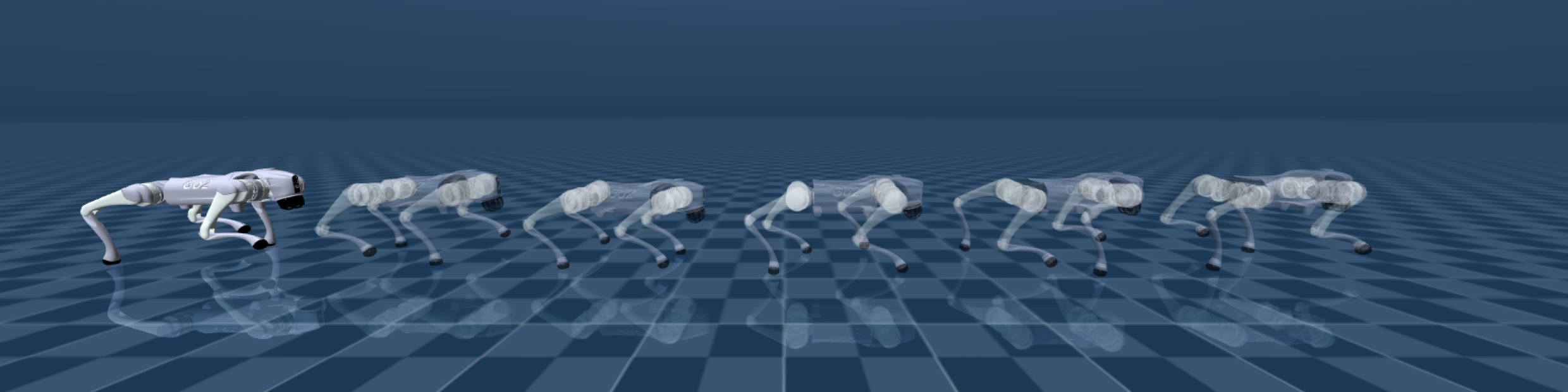}
        \label{fig:go2-walk-gait-spo}
    }

    \caption{\small Quadruped gait generated by all methods, with consecutive robot poses spaced by
    $0.7$s.}
    \label{fig:go2-walk-gait}
\end{figure}

\begin{figure}[t]
    \centering
    \begin{minipage}[t]{0.5\linewidth}
        \vspace{0pt}
        \centering
        \includegraphics[height=3.5cm]{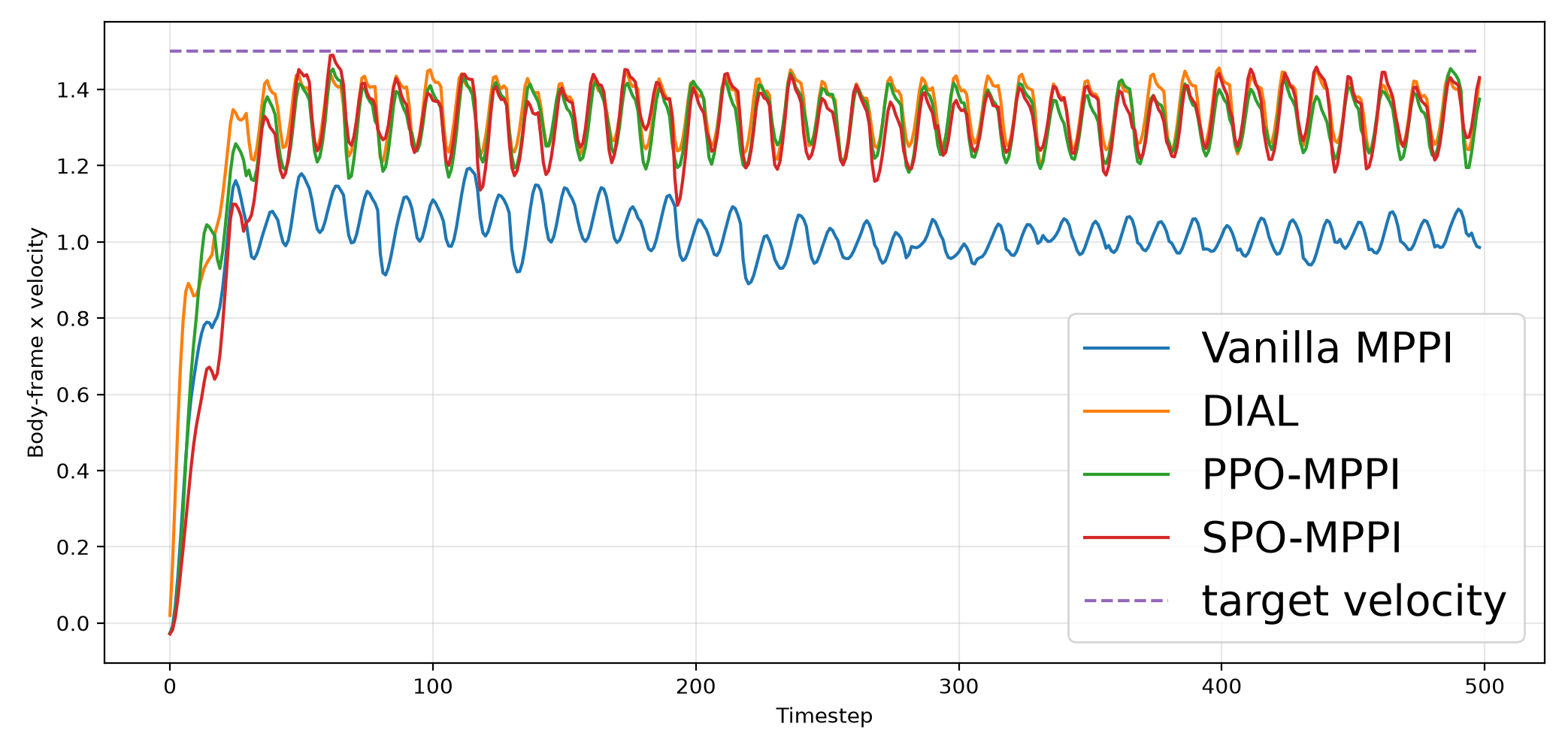}
        \captionof{figure}{Body-frame forward velocity tracking.}
        \label{fig:go2-walk-track}
    \end{minipage}
    \hfill
    \begin{minipage}[t]{0.47\linewidth}
        \vspace{5pt}
        \centering
        \small
        \setlength{\tabcolsep}{4pt}
        \begin{tabular}{lcc}
            \toprule
            & \shortstack{Avg Velocity\\(m/s)}
            & \shortstack{Computation Time\\(s) $\downarrow$} \\
            \midrule
            Vanilla-MPPI & 1.0056 & 0.0321 $\pm$ 0.0029 \\
            DIAL-MPC   & 1.3238 & 0.1373 $\pm$ 0.0015 \\
            PPO-MPPI   & 1.2887 & 0.0334 $\pm$ 0.0030 \\
            SPO-MPPI   & 1.2802 & 0.0334 $\pm$ 0.0030 \\
            \bottomrule
        \end{tabular}
        \captionof{table}{Average velocity and computation time
        for a target velocity of 1.5~m/s. 
        }
        \label{tab:go2-walk-performance}
    \end{minipage}
\end{figure}

Figure~\ref{fig:go2-walk-gait} illustrates the
gaits generated by the four methods.
Compared with Vanilla-MPPI, PPO-MPPI and SPO-MPPI
produce more regular trotting patterns with
pronounced forward leg swings.
As shown in Figure~\ref{fig:go2-walk-track}
and Table~\ref{tab:go2-walk-performance},
both methods substantially improve velocity
tracking over Vanilla-MPPI, achieving average
velocities of 1.289 and 1.280~m/s, respectively,
compared with 1.006~m/s.
Although DIAL-MPC achieves the highest average
velocity of 1.324~m/s, its computation time is
approximately $4\times$ that of the proposed
methods.
PPO-MPPI and SPO-MPPI thus achieve 
performance close to DIAL-MPC while retaining
the computational efficiency of Vanilla-MPPI,
highlighting the benefit of improving the sampling
distribution through trajectory reuse.

\subsection{Quadrotor Path Tracking}
\begin{figure}[t]
    \centering
    \subfigure[{\small Skydio X2 tracking a heart-shaped path}]{
        \begin{minipage}{0.48\linewidth}
            \centering
            \includegraphics[height=4.0cm]{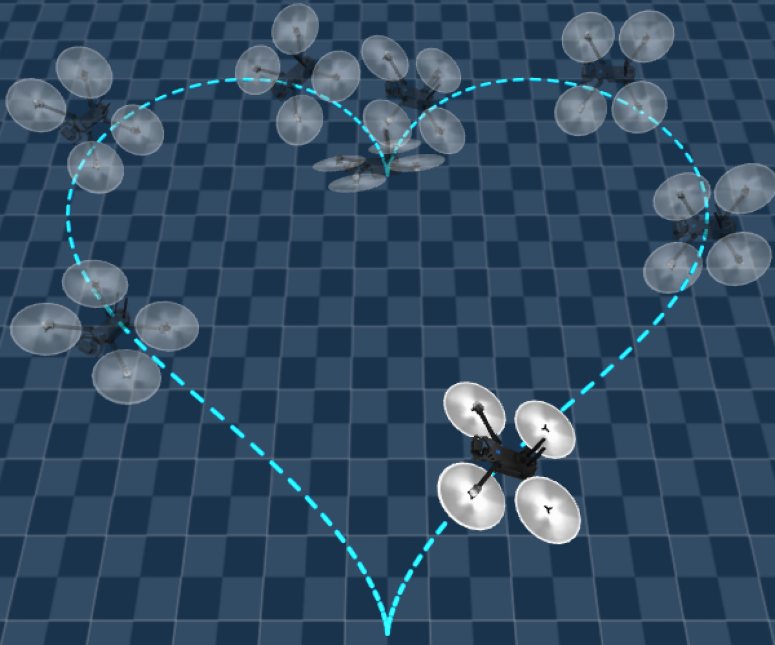}
        \end{minipage}
        \label{fig:skydio-env}
    }%
    \hfill
    \subfigure[{\small Trajectory tracking comparison}]{
        \begin{minipage}{0.48\linewidth}
            \centering
            \includegraphics[height=4.0cm]{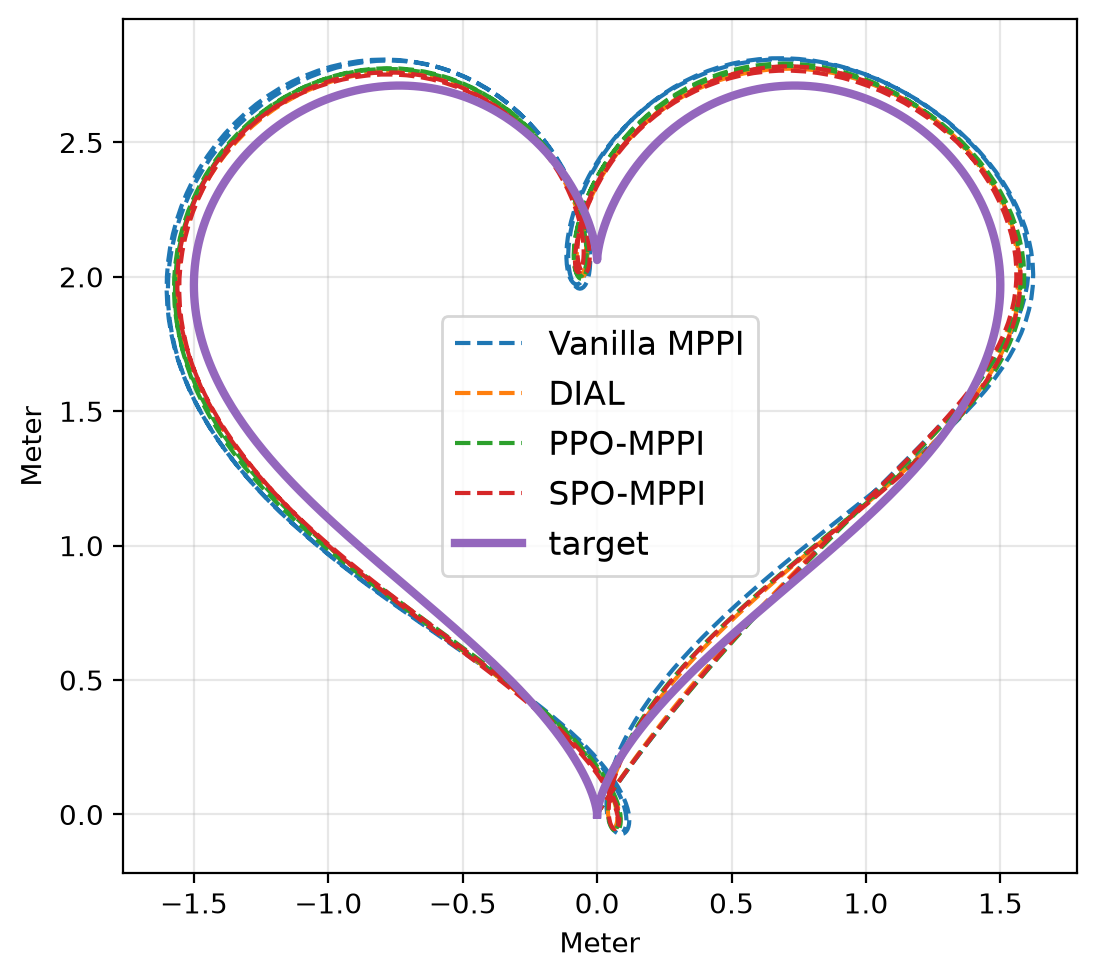}
        \end{minipage}
        \label{fig:skydio-track-2d}
    }
    \caption{\small Skydio quadrotor tracking a heart-shaped trajectory.}
    \label{fig:drone}
\end{figure}

We next evaluate the four methods on a path-tracking
task with the Skydio X2 quadrotor. Starting from
$[0,0,1]^\top\,\mathrm{m}$ with nominal hover attitude
and zero velocity, the quadrotor tracks a heart-shaped
reference trajectory of width $3\,\mathrm{m}$,
constant altitude $1\,\mathrm{m}$, and period
$5\,\mathrm{s}$. Each $10\,\mathrm{s}$ run thus
covers two periods of the reference trajectory.
Figure~\ref{fig:skydio-env} illustrates the task.
The action $\mathbf{u}\in\mathbb{R}^{4}$ specifies
additive corrections to the nominal hover thrust:
$\mathbf{u}^{\mathrm{cmd}}
=\operatorname{clip}(
\mathbf{u}^{\mathrm{hover}}+\mathbf{u},0,12)$,
where clipping is applied componentwise.
The stage cost prioritizes position tracking while
penalizing attitude deviations, angular velocity,
and control effort.


\begin{figure}[t]
    \centering
    \begin{minipage}[t]{0.5\linewidth}
        \vspace{0pt}
        \centering
        \includegraphics[height=3.5cm]{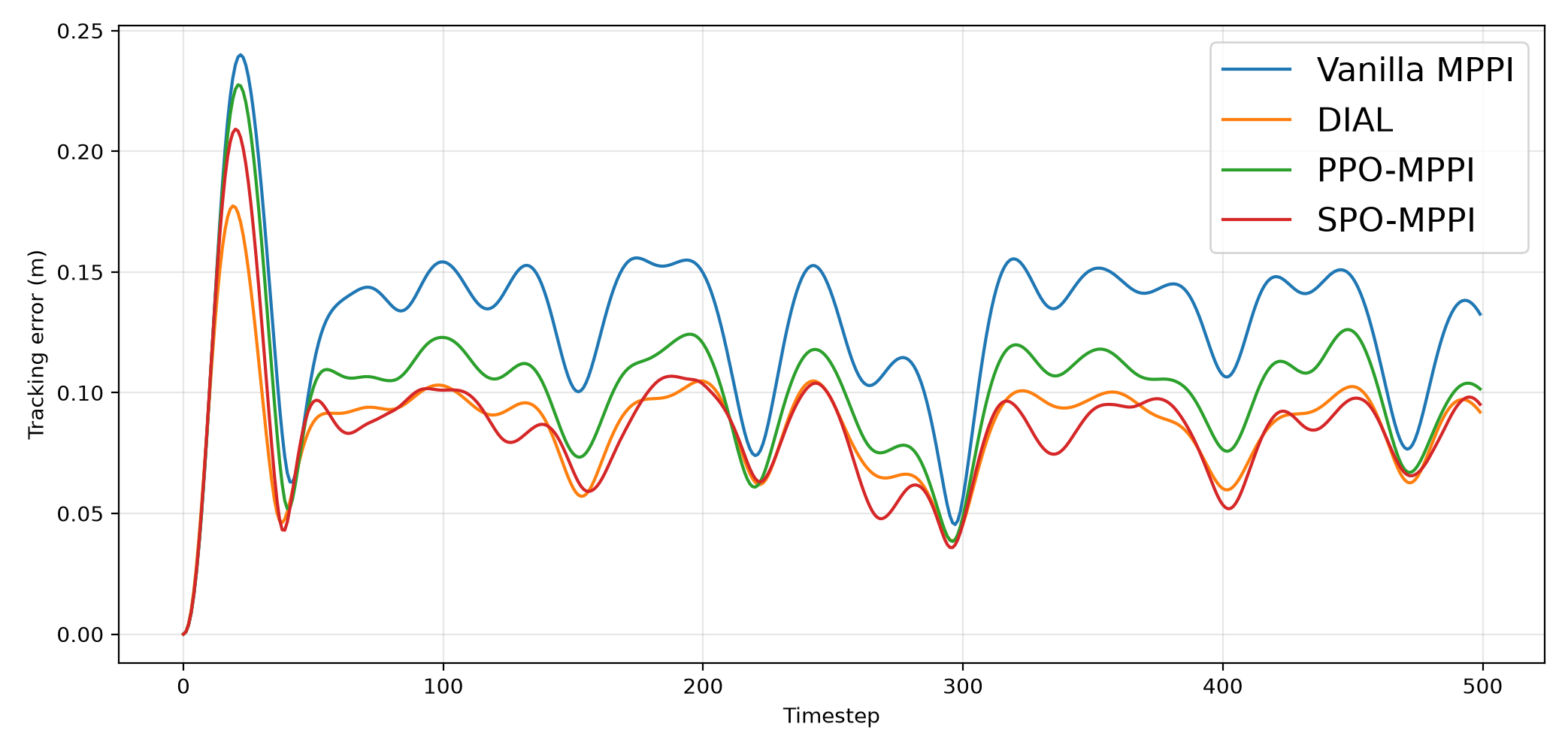}
        \captionof{figure}{\small Tracking errors of different methods over control steps.}
        \label{fig:skydio-track-error}
    \end{minipage}
    \hfill
    \begin{minipage}[t]{0.49\linewidth}
        \vspace{5pt}
        \centering
        \small
        \setlength{\tabcolsep}{4pt}
        \begin{tabular}{lcc}
            \toprule
            & \shortstack{Avg Error\\(m/s) $\downarrow$}
            & \shortstack{Computation Time\\(s) $\downarrow$} \\
            \midrule
            Vanilla-MPPI & 0.1293 & 0.0194 $\pm$ 0.0036 \\
            DIAL-MPC & 0.0866& 0.0811 $\pm$ 0.0019 \\
            PPO-MPPI & 0.1012& 0.0217 $\pm$ 0.0037 \\
            SPO-MPPI & 0.0854 & 0.0218 $\pm$ 0.0034 \\
            \bottomrule
        \end{tabular}
        \captionof{table}{ \small Mean trajectory tracking error and computation time per MPC update for the Skydio quadrotor tracking a heart-shaped trajectory
        }
        \label{tab:skydio-track-performance}
    \end{minipage}
\end{figure}
The quadrotor trajectories and tracking errors are shown in Figures~\ref{fig:skydio-track-2d} and~\ref{fig:skydio-track-error}, respectively.
Vanilla-MPPI exhibits the largest deviation from the reference. PPO-MPPI reduces the mean position error from $0.1293\,\mathrm{m}$ to $0.1012\,\mathrm{m}$, while SPO-MPPI achieves the lowest mean position error of $0.0854\,\mathrm{m}$, slightly outperforming DIAL-MPC at $0.0866\,\mathrm{m}$.
PPO-MPPI and SPO-MPPI require approximately $0.021\,\mathrm{s}$ per MPC update, compared with $0.0811\,\mathrm{s}$ for DIAL-MPC.
PPO-MPPI therefore delivers tracking performance comparable to DIAL-MPC, while SPO-MPPI surpasses it; both methods require approximately $1/4$th of DIAL-MPC's per-update computation time.

\section{Conclusion}
\label{sec:conclusion}

This paper established a connection between policy
optimization in reinforcement learning and sampling-based
MPC through the free-energy formulation of trajectory
optimization. We derived an exact performance-difference
identity and introduced PPO-MPPI and SPO-MPPI, which
enable multiple optimization updates using previously
sampled trajectories. We established population-level
monotonic improvement guarantees for both methods and
convergence to first-order stationarity for PPO-MPPI
under fixed-covariance Gaussian sampling.
Experiments on quadruped locomotion and quadrotor
tracking demonstrated that the proposed methods
achieve competitive control performance with
substantially lower computational cost than
multi-stage sampling approaches. These results
highlight the potential of proximal policy optimization
as a computationally efficient framework for
sampling-based predictive control.

\clearpage
\bibliography{l4dc2027-sample}

\clearpage

\section{Appendix}\label{sec:app}

\subsection{Complete Algorithm}
\label{app:subsec:complete algo}
\begin{algorithm2e}[h]
\caption{Proximal Policy Optimization for Sampling-Based MPC
(PPO-MPPI / SPO-MPPI)}
\label{alg:ppo_mppi}
\KwIn{Dynamics model $f$, trajectory cost $J$,
Gaussian proposal $\pi_\theta=\mathcal{N}(\theta,\Sigma)$,
temperature $\tau>0$, rollouts $N$,
proximal updates $M$, method $\mathsf{P}\in\{\mathrm{PPO},\mathrm{SPO}\}$,
parameter $\epsilon>0$, learning rate $\eta>0$}

Initialize sampling parameters $\theta_0$\;

\For{$t=0,1,2,\ldots$}{
    Observe current state $x_t$\;

    Set reference parameters $\theta_k\gets\theta_t$\;

    Sample $u^{(i)}\sim\pi_{\theta_k}$,
    $i=1,\ldots,N$\;

    Evaluate costs $J_i=J(u^{(i)};x_t)$
    using model rollouts\;

    Compute weights $w_i\gets\exp(-J_i/\tau)$
    and $\widehat Z_k\gets\frac{1}{N}\sum_{i=1}^N w_i$\;

    Compute advantages
    $\widehat A_i\gets w_i/\widehat Z_k-1$,
    $i=1,\ldots,N$\;

    Initialize $\theta^{(0)}\gets\theta_k$\;

    \For{$j=0,\ldots,M-1$}{
        Compute likelihood ratios
        $r_i\gets
        \pi_{\theta^{(j)}}(u^{(i)})
        /\pi_{\theta_k}(u^{(i)})$\;

        \eIf{$\mathsf{P}=\mathrm{PPO}$}{
            Evaluate $\widehat L_k(\theta^{(j)})
            \gets \frac{1}{N}\sum_{i=1}^N
            \ell_\epsilon(r_i,\widehat A_i)$
            using \eqref{eq:clipped_integrand}\;
        }{
            Evaluate $\widehat L_k(\theta^{(j)})
            \gets \frac{1}{N}\sum_{i=1}^N
            \left[
                r_i\widehat A_i
                -
                \frac{|\widehat A_i|}{2\epsilon}
                (r_i-1)^2
            \right]$\;
        }

        Update
        $\theta^{(j+1)}
        \gets
        \theta^{(j)}
        +
        \eta\nabla_\theta
        \widehat L_k(\theta^{(j)})$\;
    }

    Set $\theta_t^+\gets\theta^{(M)}$\;

    Apply
    $a_t=\mathbb{E}_{\pi_{\theta_t^+}}[u_0]$\;

    Shift the optimized sampling distribution
    to initialize $\theta_{t+1}$\;
}
\end{algorithm2e}

\subsection{Derivation of Optimal Sampling Distribution}
\label{app:subsec:gibbs distribution}
For any $\rho\ll\pi_\theta$, using the expression of $\rho^\star_{\theta}$ in \eqref{eq:gibbs}, the variational objective \eqref{eq:variational} can be rewritten as
\begin{equation}
    \E_\rho[J(U)]
    +
    \tau\KL(\rho\|\pi_\theta)
    =
    -\tau\log Z(\theta)
    +
    \tau\KL(\rho\|\rho_\theta^\star).
    \label{eq:gibbs_decomposition}
\end{equation}
Therefore the optimal $\rho$ that minimizes the variational objective is $\rho_\theta^\star$.

\subsection{Proof of Proposition 1}
\begin{proof}
From the definition of $A_k$, we have
\begin{align}
    \frac{Z(\theta)}{Z_k}
    &=
    \mathbb{E}_{\pi_k}
    \left[
        r_\theta(u)(1+A_k(u))
    \right]=
    1+
    \mathbb{E}_{\pi_k}
    \left[
        r_\theta(u)A_k(u)
    \right],
\end{align}
where the second equality follows from
$\mathbb{E}_{\pi_k}[r_\theta]=1$.
Taking the negative logarithm and multiplying
by $\tau$ proves the first claim. The second
claim follows from the strict monotonicity
of the logarithm.
\end{proof}

\subsection{Proof of Theorem 2}
\begin{proof}
For every $r\geq0$ and $A\in\mathbb{R}$,
the definition of the minimum gives
$
    \ell_\epsilon(r,A)\leq rA$. 
Taking expectations establishes
\eqref{eq:surrogate_lower_bound}.
At $\theta=\theta_k$, the likelihood ratio
equals one, so clipping is inactive and
\begin{equation}
    L_k^{\mathrm{clip}}(\theta_k)
    =
    L_k(\theta_k)=0.
\end{equation}
Therefore, if
$L_k^{\mathrm{clip}}(\theta_{k+1})\geq0$,
then $L_k(\theta_{k+1})\geq0$.
The improvement guarantee follows from
Proposition~\ref{prop:performance_difference}. Finally, combining
\eqref{eq:performance_difference} with
\eqref{eq:surrogate_lower_bound} and
the monotonicity of the logarithm proves
\eqref{eq:clipped_fe_upper_bound}.
\end{proof}

\subsection{Proof of Lemma 3}
\begin{proof}
The inequality $0<m\leq e^{-J(u)/\tau}\leq G$ 
implies
\begin{equation}
    m\leq Z(\theta)\leq G,
    \qquad
    J_{\min}\leq V(\theta)\leq J_{\max}.
\end{equation}
For any $r\geq 0$ and $A\in\mathbb{R}$,
the clipped integrand in
\eqref{eq:clipped_integrand} satisfies
\begin{equation}
\begin{aligned}
    0
    &\leq rA-\ell_\epsilon(r,A)\\
    &\leq
    |A|\,|r-1|\,
    \mathbf{1}_{\{|r-1|>\epsilon\}}\\
    &\leq
    \frac{|A|}{\epsilon}(r-1)^2,
\end{aligned}
\label{eq:pointwise_clipping_error}
\end{equation}
where the second bound follows because the clipping error vanishes unless $|r-1|>\epsilon$ and is at most $|A||r-1|$ otherwise, and the final bound uses $|r-1|\leq (r-1)^2/\epsilon$ on this event.
Furthermore, since
$m\leq e^{-J(u)/\tau}\leq G$
and $m\leq Z_k\leq G$, the signed
advantage \eqref{eq:exp_advantage} satisfies
\begin{equation}
    |A_k(u)|
    =
    \left|
        \frac{e^{-J(u)/\tau}}{Z_k}-1
    \right|
    \leq \frac{G}{m}.
\end{equation}
Substituting $r=r_\theta(u)$ and
$A=A_k(u)$ into
\eqref{eq:pointwise_clipping_error},
and taking expectations with respect to
$u\sim\pi_k$, and using \eqref{eq:cal_L_def}, yields
\begin{equation}
\begin{aligned}
    0
    &\leq
    \mathcal{L}_k(h)
    -
    \mathcal{L}_k^{\mathrm{clip}}(h) \leq
    \frac{G}{m\epsilon}
    \mathbb{E}_{\pi_k}
    \left[(r_\theta(u)-1)^2\right].
\end{aligned}
\end{equation}
For the Gaussian sampling family, let
$z=\Sigma^{-1/2}(u-\theta_k)$.
Under $u\sim\pi_k$, we have
$z\sim\mathcal{N}(0,I_d)$, and the
likelihood ratio takes the form
\begin{equation}
    r_\theta(u)
    =
    \exp\left(
        h^\top z-\frac12\|h\|^2
    \right).
\end{equation}
Using the moment-generating function
of the standard Gaussian distribution,
\begin{equation}
\begin{aligned}
    \mathbb{E}_{\pi_k}
    \left[(r_\theta(u)-1)^2\right]
    &=
    \mathbb{E}_{\pi_k}
    \left[r_\theta(u)^2\right]-1=
    e^{\|h\|^2}-1.
\end{aligned}
\end{equation}
Substituting this identity establishes
\eqref{eq:clipping_error}. Finally,
using $e^x-1\leq e^{a^2}x$ for
$0\leq x\leq a^2$ gives
\eqref{eq:local_clipping_bound}.
\end{proof}

\subsection{Proof of Theorem 4}
\begin{proof}
First, note that
$\mathcal{L}_k(0)=0$ and define
$
    q_k:=
    \nabla\mathcal{L}_k(0)
    =
    -\frac{1}{\tau}
    \Sigma^{1/2}\nabla V(\theta_k)
    \label{eq:surrogate_gradient}$.  Moreover, differentiating the Gaussian
likelihood ratio gives
\begin{equation}
\begin{aligned}
    \nabla^2\mathcal{L}_k(h)
    =
    \frac{1}{Z_k}
    \mathbb{E}_{\pi_k}
    \big[
        e^{-J(u)/\tau}r_h(z)
        \left(
            (z-h)(z-h)^\top-I_d
        \right)
    \big].
\end{aligned}
\end{equation}
Using $m\leq Z_k$,
$e^{-J(u)/\tau}\leq G$,
and the Gaussian change-of-measure
identity, we obtain
\begin{equation}
    \|\nabla^2\mathcal{L}_k(h)\|
    \leq M,
    \qquad
    M:=\frac{G}{m}(d+1).
    \label{eq:surrogate_smoothness}
\end{equation}
Therefore,
\begin{equation}
    \mathcal{L}_k(h)
    \geq
    q_k^\top h
    -
    \frac{M}{2}\|h\|^2.
\end{equation}
Combining this inequality with
Lemma~\ref{lem:clipping_error},
for $\|h\|\leq a$ we obtain
\begin{equation}
    \mathcal{L}_k^{\mathrm{clip}}(h)
    \geq
    q_k^\top h
    -
    \frac{H}{2}\|h\|^2,
    \label{eq:clipped_quadratic_lower}
\end{equation}
where $ H:=M+\frac{2Ge^{a^2}}{m\epsilon}$. Furthermore, the Gaussian score identity
implies
\begin{equation}
    q_k
    =
    \frac{1}{Z_k}
    \mathbb{E}_{u\sim\pi_k}
    \left[e^{-J(u)/\tau}z\right],
    \qquad
    z:=\Sigma^{-1/2}(u-\theta_k)\sim\mathcal{N}(0,I_d).
\end{equation}
Using $Z_k\geq m$, $e^{-J(u)/\tau}\leq G$, and
$\mathbb{E}[\|z\|^2]=d$, Cauchy--Schwarz yields
\begin{equation}
    \|q_k\|
    \leq \frac{G}{m}\mathbb{E}[\|z\|]
    \leq \frac{G}{m}\sqrt{\mathbb{E}[\|z\|^2]}
    = \frac{G}{m}\sqrt{d}
    =: Q.
\end{equation}
Choose
$t=\min\left\{
        \frac{1}{H},
        \frac{a}{Q}
    \right\}>0,$
then the candidate $h=tq_k$ satisfies
$\|h\|\leq a$. Substituting into
\eqref{eq:clipped_quadratic_lower}
gives
\begin{equation}
\begin{aligned}
    \mathcal{L}_k^{\mathrm{clip}}(tq_k)
    &\geq
    t\|q_k\|^2
    -
    \frac{Ht^2}{2}\|q_k\|^2 \geq
    \frac{t}{2}\|q_k\|^2.
\end{aligned}
\end{equation}
By optimality of $\theta_{k+1}$
and the lower-bound property
\eqref{eq:surrogate_lower_bound},
\begin{equation}
    L_k(\theta_{k+1})
    \geq
    L_k^{\mathrm{clip}}(\theta_{k+1})
    \geq
    \frac{t}{2}\|q_k\|^2.
    \label{eq:surrogate_sufficient_gain}
\end{equation}
The performance difference equality \eqref{eq:performance_difference} gives
\begin{equation}
\begin{aligned}
    V(\theta_k)-V(\theta_{k+1})
    =
    \tau\log\left(
        1+L_k(\theta_{k+1})
    \right).
\end{aligned}
\end{equation}
Since $L_k(\theta_{k+1})\geq0$ and
\begin{equation}
    1+L_k(\theta_{k+1})
    =
    \frac{Z(\theta_{k+1})}{Z_k}
    \leq\frac{G}{m},
\end{equation}
the inequality
$\log(1+x)\geq x/(1+x)$ for $x\geq0$
yields
\begin{equation}
\begin{aligned}
    V(\theta_k)-V(\theta_{k+1})
    &\geq
    \frac{\tau m}{G}
    L_k(\theta_{k+1}) \geq
    \frac{\tau mt}{2G}\|q_k\|^2=
    \frac{mt}{2\tau G}
    \|\nabla V(\theta_k)\|_\Sigma^2.
\end{aligned}
\end{equation}
Thus, \eqref{eq:convergence_descent}
holds with $c=\frac{mt}{2\tau G}>0$. Summing \eqref{eq:convergence_descent}
over $k=0,\ldots,K-1$ gives
\begin{equation}
\begin{aligned}
    c\sum_{k=0}^{K-1}
    \|\nabla V(\theta_k)\|_\Sigma^2
    &\leq
    V(\theta_0)-V(\theta_K)\leq
    V(\theta_0)-J_{\min}.
\end{aligned}
\end{equation}
Since $V(\theta_k)$ is nonincreasing
and bounded below, it converges.
Moreover,
\begin{equation}
    \sum_{k=0}^{\infty}
    \|\nabla V(\theta_k)\|_\Sigma^2
    <\infty,
\end{equation}
which implies
$\|\nabla V(\theta_k)\|\to0$
because $\Sigma\succ0$.
Finally, bounding the minimum squared
gradient norm by its average establishes
\eqref{eq:convergence_rate}.
\end{proof}

\subsection{Experiment Details}
For all tasks, we use a physics simulation time step of $0.01\,\mathrm{s}$ and hold each control input constant for two physics steps, resulting in a control period of $0.02\,\mathrm{s}$. 
All methods optimize six quadratic-spline knots with adjacent knots separated by five
control steps, yielding a 26-step prediction horizon of $0.52\,\mathrm{s}$.
Each sampling stage evaluates 1,024 trajectories with $\tau=0.25$. 
We run each controller for 500 control steps, corresponding to $10\,\mathrm{s}$ of simulated time. We use five trajectory-level annealing steps for DIAL-MPC. For PPO-MPPI, we set the clipping ratio to $\epsilon=0.9$; for SPO-MPPI, we set $\epsilon=20$. Both PPO-MPPI and SPO-MPPI optimize the spline knots using Adam \citep{kingma2014adam} with a learning rate of $0.02$ and 100 gradient updates per MPC iteration. All trajectory rollouts are simulated in parallel using MJX on an NVIDIA GeForce RTX 5080 GPU.
The complete Trust-Region PPO-MPPI procedure is summarized in Algorithm~\ref{alg:ppo_mppi}.

\subsubsection{Quadruped Velocity Tracking}
In this task, we use a reference trot gait with a cadence of $2\,\mathrm{Hz}$, a stance
duty ratio of $0.45$, and a maximum reference foot height of
$0.08\,\mathrm{m}$. Diagonally opposite feet share the same phase, and the
two diagonal pairs are offset by half a cycle.
The stage cost is
\begin{align*}
\ell ={}&
0.25\sum_{i=1}^{4}\left(\frac{z_i-z_i^{\mathrm{ref}}}{0.02}\right)^2
+0.75\frac{\left\|\mathbf{v}_{xy}^{b}-\mathbf{v}^{\mathrm{cmd}}\right\|_2^2}{0.25^2}
+0.5\left(\frac{\omega_z}{\pi}\right)^2 
+10\frac{\phi^2+\theta^2}{(0.2\pi)^2} \\
&+0.25\left(\frac{\psi}{0.1\pi}\right)^2
+0.25\left(\frac{z_b-0.3}{0.05}\right)^2 
+0.1\frac{\left\|\mathbf{q}^{\mathrm{cmd}}_j-\mathbf{q}_j^0\right\|_2^2}{(0.05\pi)^2}
+10^{-4}\left\|\dot{\mathbf{q}}_j\right\|_2^2,
\end{align*}
where $z_i$ and $z_i^{\mathrm{ref}}$ denote the actual and reference
heights of foot $i$; $\mathbf{v}_{xy}^{b}$ is the planar base velocity
expressed in the body frame; $\phi$, $\theta$, and $\psi$ are the base
roll, pitch, and yaw angles; $\omega_z$ is the base angular-velocity
component about its local $z$ axis; and $z_b$ is the base height.
The cost combines gait and velocity tracking with penalties on base
orientation, angular velocity, height deviation, joint displacement,
and joint velocity.

\begin{figure}[t]
    \centering

    \subfigure[PPO-MPPI]{
        \includegraphics[width=0.485\linewidth]{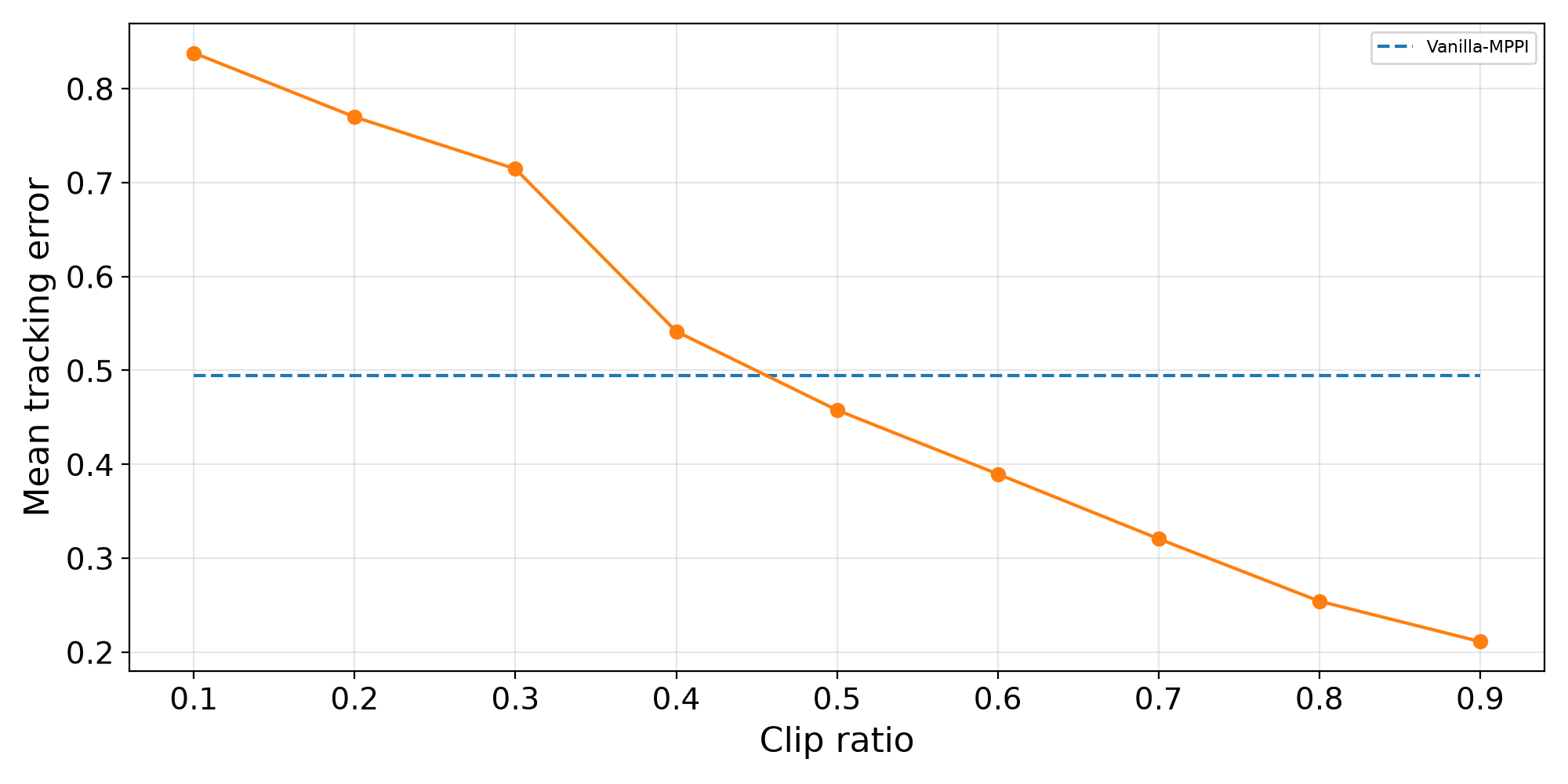}
        \label{fig:go2-walk-ppo-ablate-track-error}
    }%
    \hfill
    \subfigure[SPO-MPPI]{
        \includegraphics[width=0.485\linewidth]{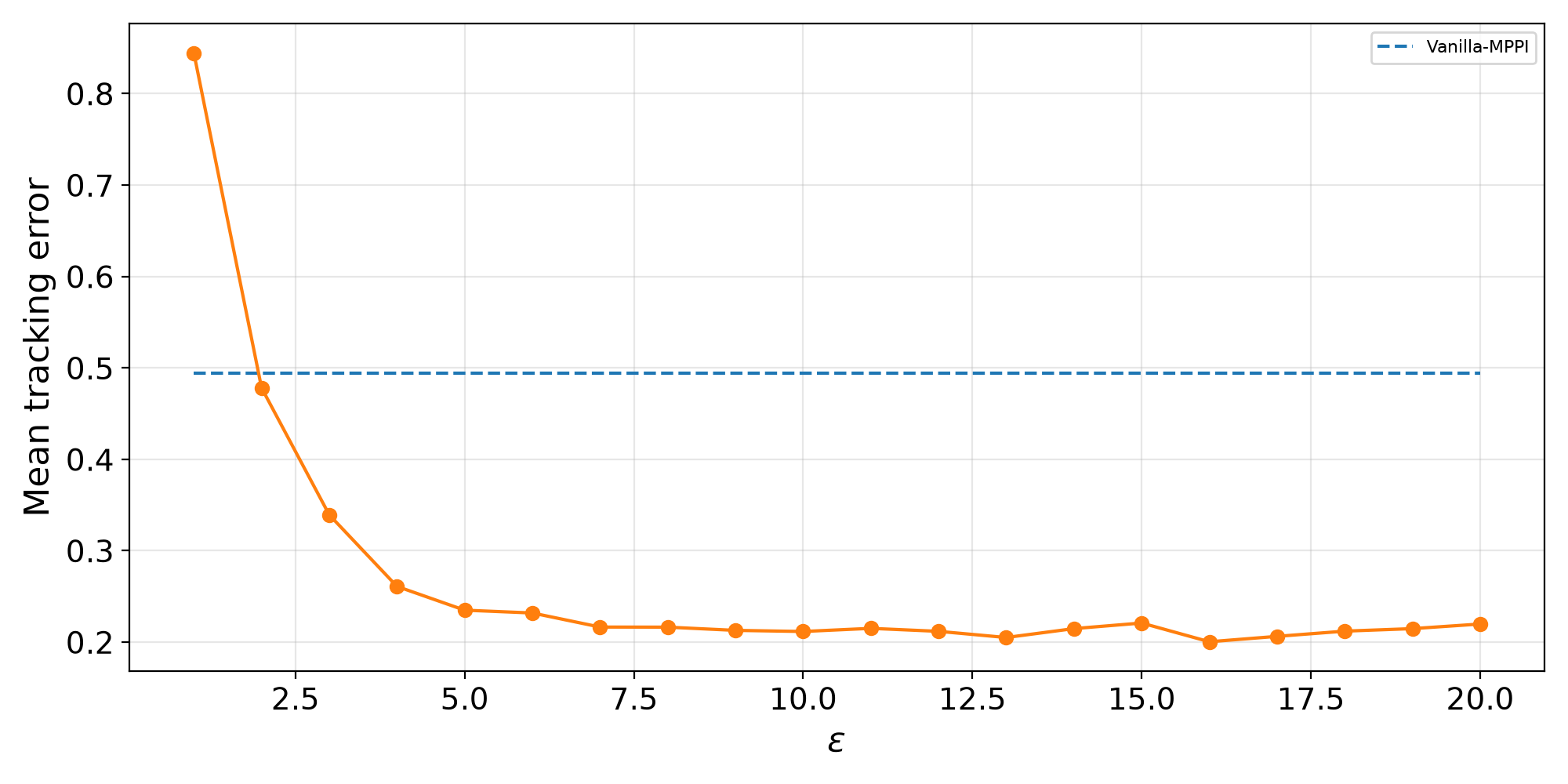}
        \label{fig:go2-walk-spo-ablate-track-error}
    }

    \caption{Ablation study on tracking error for hyperparameters of PPO-MPPI and SPO-MPPI.}
    \label{fig:go2-walk-ablation-track-error}
\end{figure}

\begin{figure}[t]
    \centering

    \subfigure[PPO-MPPI]{
        \includegraphics[width=0.485\linewidth]{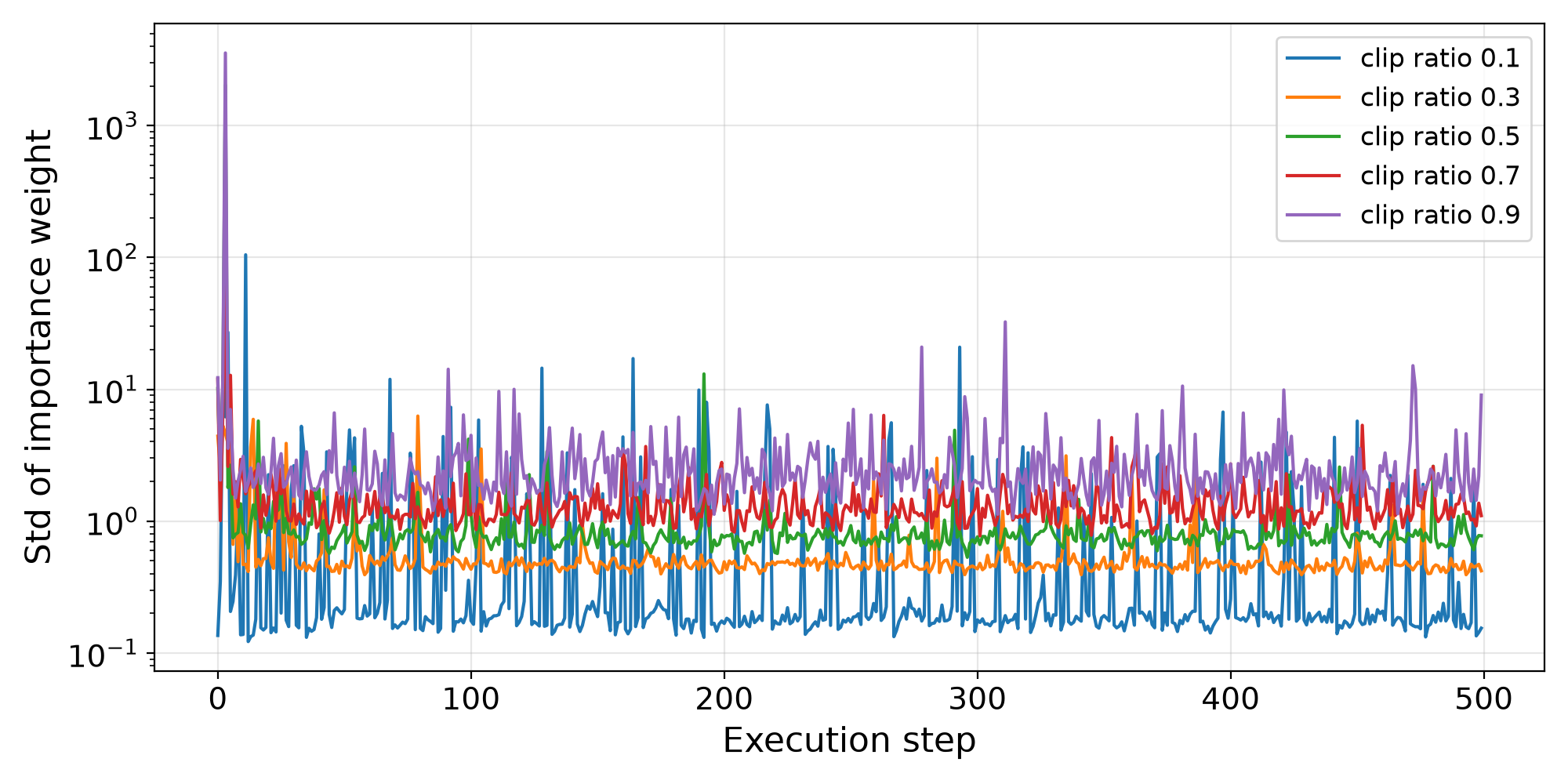}
    }%
    \hfill
    \subfigure[SPO-MPPI]{
        \includegraphics[width=0.485\linewidth]{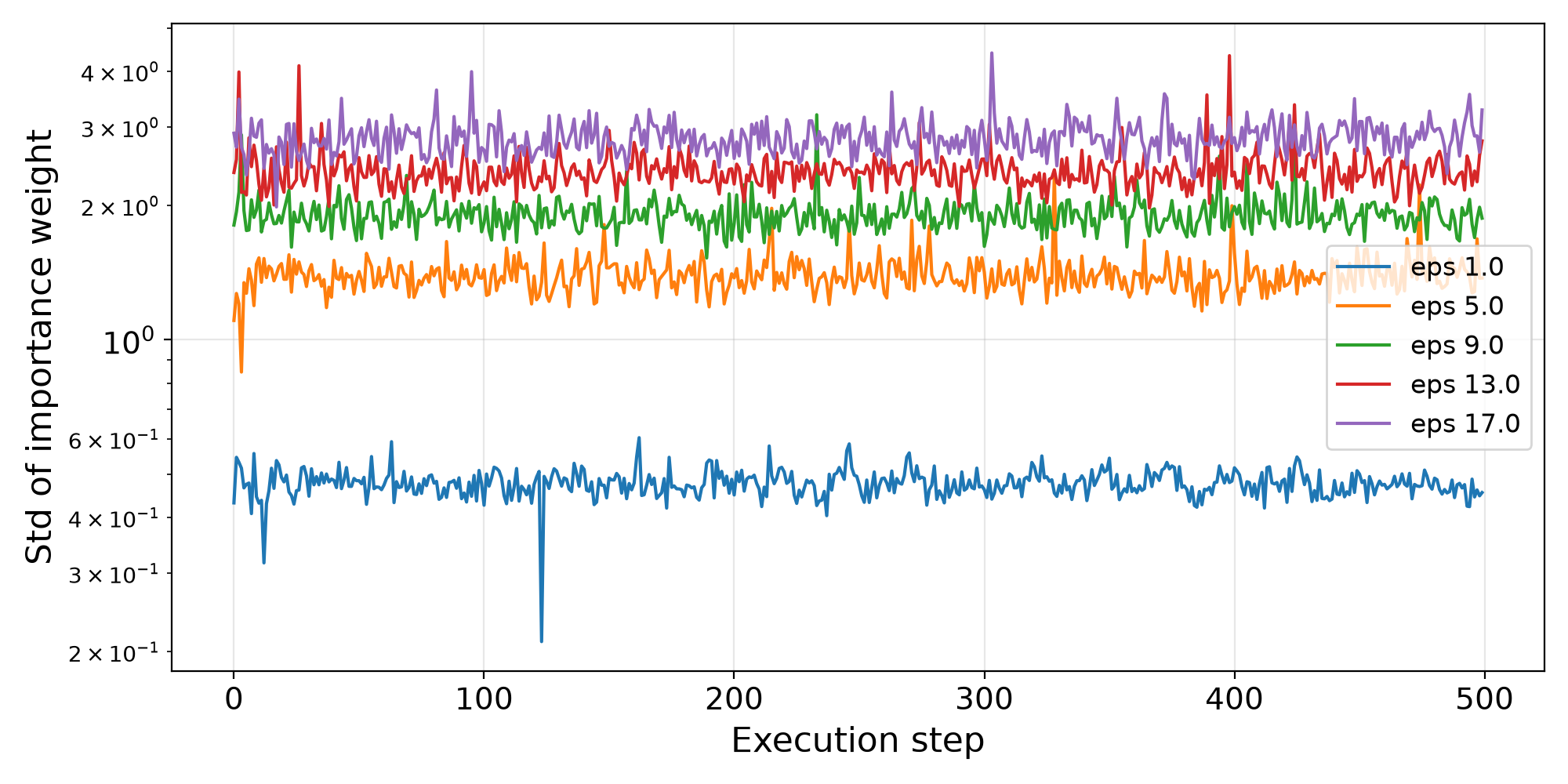}
        \label{fig:go2-walk-spo-ablate-prob-ratio}
    }

    \caption{Standard deviation of probability ratio $r_\theta(u)$ of PPO-MPPI and SPO-MPPI. $x$-axis represents execution steps}
    \label{fig:go2-walk-ablation-prob-ratio}
\end{figure}

Figure~\ref{fig:go2-walk-ablation-track-error} shows the mean velocity tracking error of PPO-MPPI and SPO-MPPI for different values of \(\epsilon\). For both methods, larger values of \(\epsilon\) permit greater distribution shifts.

Figure~\ref{fig:go2-walk-ablation-prob-ratio} shows the variability of the probability ratio $r_{\theta}(u)$ for PPO-MPPI and SPO-MPPI.
At each execution step, after maximizing the proximal surrogate, we compute the probability ratios for all sampled spline knot trajectories and report their standard deviation.
For both methods, the standard deviation increases as $\epsilon$ increases, allowing larger distribution shifts.
SPO-MPPI generally exhibits lower variability than PPO-MPPI because its regularization term directly penalizes deviations of the probability ratio from $1$, even when the ratio lies far outside the clipping range.




\subsubsection{Quadrotor Path Tracking}
In this task, a quadrotor tracks a heart-shaped reference trajectory with a
width of $3\,\mathrm{m}$, a constant altitude of $1\,\mathrm{m}$,
and a period of $5\,\mathrm{s}$. Defining $\varphi(t) = \frac{2}{5} \pi (t \bmod 5)$,
the reference position is
\[
\mathbf{p}^{\mathrm{ref}}(t)=
\begin{bmatrix}
\frac{3}{2}\sin^3\varphi(t) \\
\frac{3}{32}
\left[
17-13\cos\varphi(t)-5\cos\!\left(2\varphi(t)\right)
+2\cos\!\left(3\varphi(t)\right)
-\cos\!\left(4\varphi(t)\right)
\right] \\
1
\end{bmatrix}.
\]
Each $10\,\mathrm{s}$ run therefore covers two reference-trajectory
periods.
The stage cost is
\begin{align*}
\ell ={}&
\frac{\left\|\mathbf{p}
-\mathbf{p}^{\mathrm{ref}}(t)\right\|_2^2}{0.025^2}
+10\frac{\phi^2+\theta^2}{(0.1\pi)^2}
+\frac{\omega_z^2}{(0.1\pi)^2}
+
\frac{\omega_x^2+\omega_y^2}{(0.25\pi)^2}
+2.5
\left\|\mathbf{u}^{\mathrm{cmd}}
-\mathbf{u}^{\mathrm{hover}}\right\|_2^2,
\end{align*}
where $\mathbf{p}$ is the quadrotor position; $\phi$ and $\theta$ are its roll
and pitch angles; $\boldsymbol{\omega}=[\omega_x,\omega_y,\omega_z]^\top$ is
its angular velocity; and $\mathbf{u}^{\mathrm{hover}}$ is the nominal hover
thrust. The cost therefore prioritizes position tracking while regularizing
attitude, angular velocity, and control effort.

\end{document}